\documentclass[cameraready]{Interspeech}

\title{Online Predictive Coding for Dual-Mode Self-Supervised Speech Models}

\author[affiliation={1}]{Keita}{Goto}
\author[affiliation={1}]{Takashi}{Maekaku}
\author[affiliation={1}]{Jin}{Sakuma}
\author[affiliation={2}]{Jinchuan}{Tian}
\author[affiliation={1}]{Yusuke}{Shinohara}
\author[affiliation={2}]{Shinji}{Watanabe}
\address{
    $^1$ LY Corporation, Tokyo, Japan \\
    $^2$ Carnegie Mellon University, PA, USA
}

\email{keitgoto@lycorp.co.jp, tmaekaku@lycorp.co.jp, jsakuma@lycorp.co.jp, jinchuat@andrew.cmu.edu, yusshino@lycorp.co.jp, shinjiw@ieee.org}


\usepackage{comment,booktabs,multirow,subcaption}

\newcommand{\vect}[1]{\bm{#1}}
\newcommand{\matr}[1]{\bm{#1}}
\newcommand{\online}{\mathrm{on}}
\newcommand{\offline}{\mathrm{off}}

\newcommand{\X}{\matr{X}}

\newcommand{\LA}{\matr{L}}
\newcommand{\Reg}{\matr{R}}
\newcommand{\C}{\matr{C}}

\newcommand{\Nc}{N_\mathrm{c}}
\newcommand{\Nl}{N_\mathrm{l}}
\newcommand{\Nr}{N_\mathrm{r}}
\newcommand{\Nf}{N_\mathrm{f}}

\begin{document}

\maketitle

% the abstract here must exactly match the abstract entered into the paper submission system
\begin{abstract}
% 1000 characters. ASCII characters only. No citations.
Dual-mode self-supervised speech models are pre-trained to handle streaming and non-streaming conditions simultaneously. However, their attention is computed over different context ranges, which often makes optimization difficult. In previous work, we proposed \textit{online registers}, additional tokens intended to compensate for missing future context in streaming mode, but the gains remained limited. To address these issues, we introduce two improvements for robust dual-mode pre-training: (1) Online Predictive Coding (OPC), which regularizes the registers through multi-step future prediction, and (2) Dual-mode Layer Normalization, which stabilizes optimization. We fine-tune the proposed dual-mode self-supervised speech models for speech recognition on LibriSpeech and WSJ. Results show that OPC consistently reduces the online--offline performance gap; at 160~ms latency on LibriSpeech, word error rates improve from 3.65\% to 3.40\% on \textit{test-clean} and from 10.15\% to 9.65\% on \textit{test-other}.
\end{abstract}

\section{Introduction}

Self-supervised speech models (S3Ms)~\cite{speech-ssl} have become a foundational paradigm for a wide range of speech processing tasks. By pre-training on large-scale unlabeled speech, they achieve strong performance across diverse downstream tasks, as evidenced by SUPERB~\cite{superb} and multilingual speech recognition benchmarks~\cite{xeus,xls-r,usm}. However, most leading S3Ms, such as wav2vec~2.0~\cite{wav2vec}, HuBERT~\cite{hubert}, and BEST-RQ~\cite{best-rq}, are trained under \textit{offline} (non-streaming) conditions. As a result, their performance often degrades when deployed in real-time \textit{online} (streaming) applications, where future context is unavailable.

To enable streaming capability, several adaptation strategies for S3Ms have been explored. Distillation-based approaches~\cite{transducer-w2v2,streaming-w2v2,distil-w2v2} train a streaming model to mimic a strong non-streaming teacher. While effective, these methods typically require complex training pipelines. Dual-mode frameworks~\cite{wav2vec-s,ufo2,durep} offer a more direct solution by jointly pre-training in online and offline modes within a single model. However, sharing parameters across the two modes introduces optimization challenges, and such models often underperform compared to single-mode counterparts.

To reduce the discrepancy between online and offline modes, our prior work introduced online registers~\cite{online-registers}, learnable tokens appended to each chunk in online mode. These registers provide additional representational capacity to compensate for missing future context, thereby mitigating the attention mismatch between modes. That work further incorporated a future prediction objective within the self-supervised learning framework, pre-training each register to directly predict specific future frames so as to explicitly encode future context. While this constraint encouraged the registers to encode future information, the overall gains remained limited.

In this paper, we propose an improved pre-training framework for dual-mode S3Ms using online registers to reduce the gap between online and offline modes. The framework consists of two main modifications. First, we extend the original future prediction constraint into a more flexible \textbf{Online Predictive Coding (OPC)} framework inspired by Contrastive Predictive Coding~\cite{cpc}. OPC is jointly optimized with the wav2vec~2.0 framework, encouraging the online pathway to learn representations that are predictive of future context. Second, we adopt \textbf{Dual-mode Layer Normalization}, originally proposed for automatic speech recognition (ASR), in SSL training by maintaining separate LayerNorm~\cite{layernorm} parameters for online and offline modes to mitigate mode-specific distribution shifts.

These modifications stabilize dual-mode pre-training and enhance robustness in both online and offline downstream tasks without increasing algorithmic latency. On the LibriSpeech ASR benchmark at 160~ms latency, OPC reduces online word error rate (WER) from 3.65\% to 3.40\% on \textit{test-clean} and from 10.15\% to 9.65\% on \textit{test-other}. In parallel, offline WER improves from 2.73\% to 2.64\% and from 6.63\% to 6.41\% on \textit{test-clean} and \textit{test-other}, respectively, helping narrow the online--offline performance gap.

\section{Related Work}

\noindent
\textbf{Dual-mode architectures for ASR.}
Dual-mode models support both online and offline inference within a single architecture and have been studied in automatic speech recognition (ASR)~\cite{dual-mode-original,dual-mode-transformer,conformer-dual}. They are typically trained by jointly optimizing online and offline pathways with a shared encoder, often combined with Dynamic Chunk Training~\cite{dynamic-chunk-training} to support variable-latency conditions. 

However, these models are specialized for the ASR task, and information such as speaker identity or emotion may not be inherently embedded in the model representations. By focusing on \textit{dual-mode self-supervised models}, our approach aims for deployment across a wide range of streaming and non-streaming applications beyond ASR.

\noindent
\textbf{Dual-mode self-supervised learning.}
With the growing demand for diverse applications, dual-mode strategies have been extended to self-supervised speech models (S3Ms). Wav2vec-S~\cite{wav2vec-s} and UFO2~\cite{ufo2} apply dual-mode pre-training following the wav2vec 2.0 pre-training scheme, while DuRep~\cite{durep} focuses on ASR-oriented multi-stage pre-training.

In previous work, we proposed online registers~\cite{online-registers}, learnable tokens appended to each chunk in online mode to mitigate attention mismatch caused by missing future context. Although these registers alleviate the discrepancy between online and offline pathways, they were not explicitly regularized to encode future information, and the overall gains remained limited. 

Motivated by the need to make online registers capture more information about the unseen future context, we draw inspiration from Contrastive Predictive Coding (CPC)~\cite{cpc}, which learns contextual representations by predicting future latent states from past representations. We extend this predictive-learning principle to dual-mode S3Ms and develop \textit{Online Predictive Coding} (OPC) built upon online registers. Unlike CPC, OPC is jointly optimized with the masked prediction objective of wav2vec~2.0 during pre-training, enabling the model to acquire representations that leverage both preceding and succeeding context.

\noindent
\textbf{Normalization and parameter sharing.}
Another important design choice concerns parameter sharing in normalization layers. The original dual-mode ASR study~\cite{dual-mode-original} advocates separating normalization parameters for each mode, while other approaches~\cite{durep,conformer-dual} maintain shared normalization across modes. When additional learnable tokens, such as online registers, are introduced, shared normalization may amplify statistical discrepancies between pathways. We revisit this design choice and adopt \textit{Dual-mode Layer Normalization} to stabilize dual-mode optimization.

\section{Methods}
\label{sec:proposed}

\subsection{Dual-mode Transformer with Online Registers}
\label{sec:dm_register}

Our encoder is based on wav2vec 2.0~\cite{wav2vec} and is pre-trained in a dual-mode manner. Let $\X=(\vect{x}_1,\dots,\vect{x}_T)$ denote the latent features extracted by a convolutional feature encoder, where $\vect{x}_t\in\mathbb{R}^d$ and $T$ is the number of feature frames. For online mode, we partition $\X$ into chunks of size $\Nc$ with an optional lookahead of $\Nl$ frames. For the $i$-th chunk, the chunk $\C_i$ and lookahead $\LA_i$ are defined as
\begin{equation}
\C_i=\X_{(i-1)\Nc+1:i\Nc},\quad
\LA_i=\X_{i\Nc+1:i\Nc+\Nl}.
\end{equation}
We append $\Nr$ learnable online registers to each chunk: $\Reg_i=(\vect{r}_{1},\dots,\vect{r}_{\Nr})$, where the register embeddings $\{\vect{r}_{m}\}_{m=1}^{\Nr}$ are shared across all chunks.

We denote the Transformer encoder by $f(\cdot)$. In online mode, an attention mask $\matr{M}$ masks out positions beyond the current chunk and lookahead\footnote{In this paper, online mode masks only future context; past context remains fully visible.}. During training, we concatenate $\C_i$, $\LA_i$, and $\Reg_i$ over all chunks as $\C$, $\LA$, and $\Reg$, respectively, and encode them in parallel using $\matr{M}$ to emulate streaming constraints~\cite{online-registers}:
\begin{equation}
\label{eq:dm_forward}
\begin{aligned}
\left(\hat{\C}^{\online},\hat{\LA}^{\online},\hat{\Reg}^{\online}\right) &= f\big((\C,\LA,\Reg); \matr{M}\big), \\
\hat{\X}^{\offline} &= f(\X).
\end{aligned}
\end{equation}
At inference time, online representations are extracted in a chunk-by-chunk manner without attention masks. 
Online registers introduce only a marginal increase in memory and computation, without affecting algorithmic latency.

\subsection{Why Online--Offline Attention Mismatch Happens}
\label{sec:mismatch}

\begin{figure}[t]
  \centering
  \includegraphics[width=\linewidth]{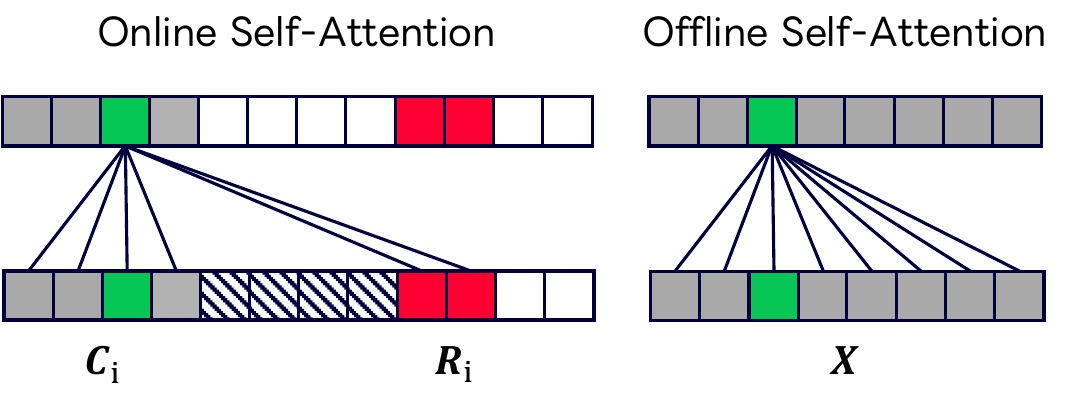}
  \vspace{-12pt}
  \caption{Online vs. offline self-attention visibility. The figure illustrates the case where the chunk size is $\Nc=4$, with $\Nr=2$ registers per chunk and no lookahead ($\Nl=0$). Offline attention attends to all frames in $\X$, whereas online attention is restricted to the chunk $\C_i$ and its online registers $\Reg_i$.}
  \label{fig:attention}
\end{figure}

The key difference between online and offline modes lies in the context available to self-attention. Offline self-attention can attend to the entire utterance, whereas online self-attention is restricted to the current chunk and lookahead. Since dual-mode models share the same parameters across both modes, this mismatch makes optimization challenging.

Online registers mitigate this issue by introducing \textit{extra learnable slots} in online mode, providing a surrogate pathway for the missing future context. As illustrated in Figure~\ref{fig:attention}, offline attention can reference all frames, whereas online attention is restricted to the past and current chunks along with the appended registers. If the registers encode sufficient future information, the attention behaviors of the two modes become more aligned. The remaining question is how to ensure that the registers stably encode future context; we address this with \textit{Online Predictive Coding}.

\subsection{Online Predictive Coding}
\label{sec:opc}

\begin{figure}[t]
  \centering
  \includegraphics[width=\linewidth]{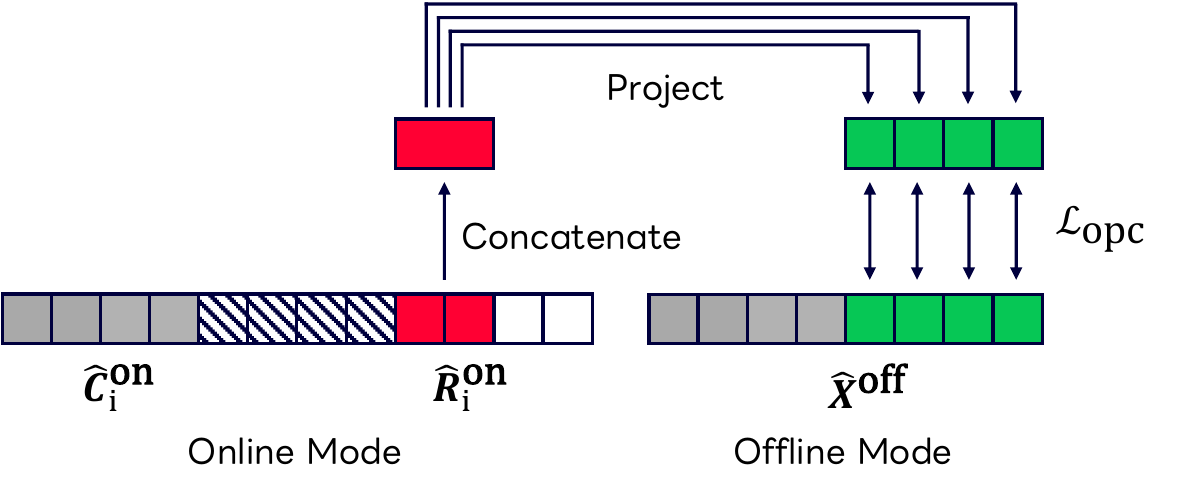}
  \vspace{-18pt}
  \caption{Overview of Online Predictive Coding (OPC) for the case of $\Nc=4$, $\Nr=2$ registers per chunk, no lookahead ($\Nl=0$), and $\Nf=4$ future prediction steps. For the $i$-th chunk, the online registers jointly predict four future offline representations at subsequent time steps. The OPC loss is computed over all chunks.}
  \label{fig:opc-overview}
\end{figure}

Online registers act as surrogates for the unseen future, but the key challenge is to make them robustly encode future information. To address this, we propose \textbf{Online Predictive Coding (OPC)}, a future prediction objective that trains the online registers to jointly predict multiple future offline representations at arbitrary offsets. Similar to CPC~\cite{cpc} and NEST-RQ~\cite{nest-rq}, OPC encourages the registers to predict future context. However, by jointly optimizing OPC with wav2vec~2.0's masked prediction objective, our model can still leverage bidirectional context during pre-training and thus acquire stronger contextual representations. Unlike these previous methods, our approach performs multi-step feature prediction from register representations at each frame, since the model can access future frames within each chunk.

Figure~\ref{fig:opc-overview} illustrates an overview of OPC.Let $\hat{\Reg}_i^\online=(\hat{\vect{r}}_{i,1},\dots,\hat{\vect{r}}_{i,\Nr})$ denote the output representations of the registers associated with chunk $i$, as defined in Equation~\ref{eq:dm_forward}. We concatenate them and predict $\Nf$ future offline representations using linear projections $\matr{W}_j\in\mathbb{R}^{d\Nr \times d}$:
\begin{equation}
\label{eq:opc_pred}
\begin{aligned}
\hat{\vect{f}}_{i,j}
&= \matr{W}_j^\top
\begin{bmatrix}
\hat{\vect{r}}_{i,1}\\[-0.2ex]
\vdots\\[-0.2ex]
\hat{\vect{r}}_{i,\Nr}
\end{bmatrix}
\in \mathbb{R}^d,\\
\vect{f}_{i,j}
&= \hat{\vect{x}}^{\offline}_{\,i\Nc+\Nl+j},
\quad j=1,\dots,\Nf.
\end{aligned}
\end{equation}
For each chunk, we predict $\Nf$ future representations from the offline pathway and define $\mathcal{L}_{\mathrm{opc}}$ as the objective that minimizes the cosine distance between vectors:
\begin{equation}
\label{eq:opc_loss}
\mathcal{L}_{\mathrm{opc}}
=
\sum_{i=1}^{\lfloor T/\Nc \rfloor + 1}\sum_{j=1}^{\Nf}
\left(1-\cos \left( \hat{\vect{f}}_{i,j},\operatorname{SG} \left( \vect{f}_{i,j} \right) \right) \right).
\end{equation}
We stop the gradient through the offline target using $\mathrm{SG}(\cdot)$ to avoid collapsing the offline pathway. This objective makes the registers predictive: to minimize $\mathcal{L}_{\mathrm{opc}}$, the online registers must encode information useful for reconstructing future offline representations. Consequently, OPC encourages the registers to approximate the contribution of the unseen future frames, as illustrated in Figure~\ref{fig:attention}.

We jointly optimize the OPC loss with the wav2vec 2.0 objectives for online and offline modes ($\mathcal{L}^\online$ and $\mathcal{L}^\offline$) and the codebook diversity loss $\mathcal{L}_{\mathrm{d}}$:
\begin{equation}
\label{eq:total_loss}
\mathcal{L}=
\frac{1}{2}\left(\mathcal{L}^\online+\mathcal{L}^\offline\right)
+w_{\mathrm{d}}\mathcal{L}_{\mathrm{d}}
+w_{\mathrm{opc}}\mathcal{L}_{\mathrm{opc}},
\end{equation}
where $w_{\mathrm{d}}$ and $w_{\mathrm{opc}}$ are hyperparameters.

\subsection{Dual-mode Layer Normalization}
\label{sec:dm_ln}

Dual-mode training inherently induces mode-dependent feature statistics. As analyzed in Section~\ref{sec:mismatch}, in dual-mode attention, different statistics are computed for the online and offline pathway. Furthermore, the introduction of online registers exacerbates this discrepancy; as these learnable tokens are explicitly regularized by OPC to compensate for the missing future frames, their activation patterns and norms can differ significantly from those of standard speech frames.

To mitigate interference between modes, we adopt \textbf{Dual-mode Layer Normalization}~\cite{dual-mode-original}: each LayerNorm has separate affine parameters ($\vect{\gamma}$ and $\vect{\beta}$) for offline and online modes, while all other weights are shared:
\begin{equation}
\left( \vect{\gamma},\vect{\beta} \right)=
\begin{cases}
\left( \vect{\gamma}^{\offline},\vect{\beta}^{\offline} \right) & \text{for offline mode},\\
\left( \vect{\gamma}^{\online},\vect{\beta}^{\online} \right) & \text{for online mode}.
\end{cases}
\end{equation}
This simple decoupling stabilizes optimization under distribution shifts and improves robustness across latency conditions.

\section{Experiments}

\subsection{Experimental Settings}

Unless otherwise stated, we follow the official Fairseq~\cite{fairseq} wav2vec 2.0 configurations as our baseline and introduce only the modifications described in Section~\ref{sec:proposed}.

\noindent
\textbf{Datasets.}
Pre-training was conducted on the 960-hour LibriSpeech corpus~\cite{librispeech} without transcriptions. For ASR fine-tuning, we used LibriSpeech 960h and report WER on \textit{test-clean} and \textit{test-other}. For cross-domain evaluation, we additionally fine-tuned and evaluated on the Wall Street Journal (WSJ)~\cite{wsj} corpus using \textit{train\_si284} for training, \textit{test\_dev92} for validation, and \textit{test\_eval92} and \textit{test\_eval93} for testing.

\noindent
\textbf{Pre-training.}
We follow the official Fairseq wav2vec 2.0 BASE configuration for LibriSpeech 960h. Our model is pre-trained in a dual-mode manner as formulated in Equation~\ref{eq:total_loss} with shared parameters except for LayerNorm. Streaming behavior is simulated using chunk-based attention masks~\cite{streaming-tranformer} together with Dynamic Chunk Training (DCT)~\cite{dynamic-chunk-training}, where the chunk size and lookahead are sampled as $\Nc \sim \mathcal{U}(2,32)$ and $\Nl \sim \mathcal{U}(0,\Nc)$, where $\mathcal{U}$ is the uniform distribution. In our experiment, the number of online registers $\Nr$ is fixed at 1. The weight parameters in Equation~\ref{eq:total_loss} are set to $w_\mathrm{d}=0.1$ and $w_\mathrm{opc}=0.1$.

Pre-training is conducted for 100k steps using 16 NVIDIA H200 GPUs. The learning rate is linearly warmed up to $1 \times 10^{-4}$ over the first 8k steps and then linearly decayed to zero for the remaining steps. The batch size corresponds to 350 seconds of audio per GPU. All parameters, including the online registers' embeddings, are optimized using Adam~\cite{adam}.

Following wav2vec-S~\cite{wav2vec-s}, we removed relative positional encodings and instead applied sinusoidal positional encodings. Models were initialized from the official checkpoint trained on LibriSpeech 960h. The parameters for Dual-mode Layer Normalization are initialized using the parameters from the checkpoint for both modes.

\noindent
\textbf{Fine-tuning and Decoding.}
For LibriSpeech, we adopt the official Fairseq wav2vec 2.0 BASE 960h fine-tuning configuration, while for WSJ we follow the Fairseq LibriSpeech 100h setup. We replace the pre-training projection head with a linear layer and optimize the average Connectionist Temporal Classification (CTC)~\cite{ctc} loss over online and offline modes. Dynamic Chunk Training is also applied during fine-tuning to support arbitrary chunk sizes. Fine-tuning is performed for 320k steps on 8 NVIDIA H200 GPUs. The batch size corresponds to 200 seconds of audio per GPU.

For LibriSpeech decoding, we use the Flashlight beam search decoder~\cite{flashlight} with the officially released 4-gram language model and a beam size of 50. For WSJ, we use the same decoding framework with a 4-gram language model trained on the WSJ text data. 

\subsection{Main Results}

Table~\ref{tab:main-results} summarizes ASR performance on LibriSpeech and WSJ in both online and offline settings, using the dual-mode model without online registers as the baseline. We first consider the LibriSpeech streaming condition with a chunk size of $\Nc=8$ and zero lookahead ($\Nl=0$), which corresponds to a low-latency setting (160\,ms). Under this condition, adding online registers yields relative WER reductions of 4.1\% on \textit{test-clean} and 3.4\% on \textit{test-other} in online mode. Incorporating Online Predictive Coding (OPC) further improves streaming recognition, achieving relative WER reductions of 6.8\% on \textit{test-clean} and 4.9\% on \textit{test-other}. These results support our hypothesis that predictive supervision encourages the registers to store information that compensates for missing future context. This narrows the offline--online attention mismatch as mentioned in Section~\ref{sec:mismatch} and improves low-latency streaming ASR.

On WSJ \textit{eval93}, OPC slightly degrades offline performance (a 1.2\% relative increase in WER), suggesting that the auxiliary future-prediction task can introduce domain-specific bias when the target domain differs from the pre-training distribution. Nevertheless, it still outperforms the dual-mode baseline.

\begin{table*}[tbp]
    \centering
    \caption{Word error rate (WER, \%) of the dual-mode baseline and our proposed variants after ASR fine-tuning, evaluated on LibriSpeech (\textit{test-clean}, \textit{test-other}) and WSJ (\textit{eval92}, \textit{eval93}). Online decoding uses a \textbf{160~ms} chunk size ($N_c=8$) with no look-ahead ($N_l=0$). For Online Predictive Coding, we show the best-performing setting with the number of predicted future frames $N_f=4$.}
    \label{tab:main-results}
    \begin{tabular}{@{}lcccccccc@{}}
        \toprule
        \multirow{2}{*}{Pre-train Method}  & \multicolumn{2}{c}{test-clean}   & \multicolumn{2}{c}{test-other} & \multicolumn{2}{c}{eval92}   & \multicolumn{2}{c}{eval93}      \\ \cmidrule(l){2-9}
                                           & Offline          & Online        & Offline       & Online         & Offline       & Online        & Offline        & Online        \\ \midrule
        Dual-mode Only                     & 2.73             & 3.65          & 6.63          & 10.15          & 7.24          & 8.99          & 10.08          & 8.12          \\
        \quad w/ Online Registers          & 2.70             & 3.50          & 6.52          & 9.80           & 7.14          & 8.65          & \textbf{10.05} & \textbf{7.60} \\
        \qquad w/ Online Predictive Coding & \textbf{2.64}    & \textbf{3.40} & \textbf{6.41} & \textbf{9.65}  & \textbf{6.94} & \textbf{8.13} & 10.20          & 7.86          \\ \bottomrule
    \end{tabular}
    \vspace{-5mm}
\end{table*}

Table~\ref{tab:other-methods} compares our approach with representative self-supervised baselines under a comparable latency setting: a 640\,ms chunk size with zero lookahead. While offline-only wav2vec~2.0~\cite{wav2vec} does not support online inference, dual-mode approaches enable both online and offline recognition within a single model. Our method achieves offline performance comparable to wav2vec~2.0 and lower online WER than UFO2~\cite{ufo2} on both \textit{test-clean} and \textit{test-other}.\footnote{We note that this is not a strictly controlled comparison, since UFO2 uses attention rescoring with a Transformer decoder, whereas our model uses an $n$-gram language model.} These results indicate that improving dual-mode consistency can enhance online recognition without sacrificing offline accuracy.

\begin{table}[tbp]
    \centering
    \caption{Word error rate (WER, \%) on LibriSpeech \textit{test-clean} and \textit{test-other} for prior methods and our model. For fair comparison, all online results are decoded with a \textbf{640~ms} chunk size (for our model, $N_c=32$) and no look-ahead ($N_l=0$). Our model additionally uses Online Predictive Coding with $N_f=4$ predicted future frames. Values are rounded to one decimal place for consistency with results reported in prior work.}
    \label{tab:other-methods}
    \begin{tabular}{@{}lcccc@{}}
    \toprule
    \multirow{2}{*}{Method}                  & \multicolumn{2}{c}{test-clean} & \multicolumn{2}{c}{test-other} \\ \cmidrule(l){2-5} 
                                             & Offline      & Online          & Offline      & Online          \\ \midrule
    wav2vec 2.0~\cite{wav2vec}               & \textbf{2.6} & -               & \textbf{6.1} & -               \\
    UFO2~\cite{ufo2}                         & 3.0          & 3.8             & 7.1          & 9.4             \\ \midrule
    Ours                                     & \textbf{2.6} & \textbf{3.1}    & 6.4          & \textbf{8.3}    \\ \bottomrule
\end{tabular}
\end{table}

\subsection{Analysis}

\begin{figure}[tbp]
    \centering
    
    \begin{subfigure}[b]{0.23\textwidth}
        \centering
        \includegraphics[width=\linewidth]{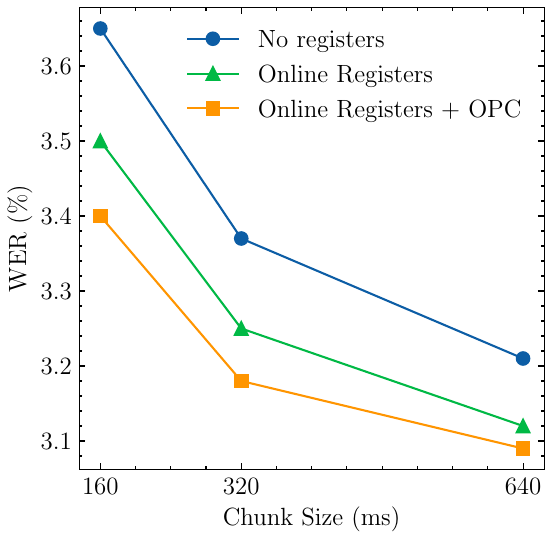}
        \caption{}
        \label{fig:1a}
    \end{subfigure}
    \begin{subfigure}[b]{0.23\textwidth}
        \centering
        \includegraphics[width=\linewidth]{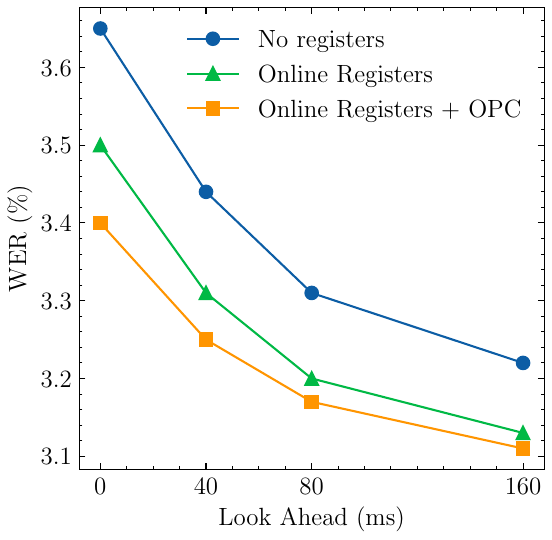}
        \caption{}
        \label{fig:1b}
    \end{subfigure}
    \vspace{-4pt}
    \caption{Word error rate (WER, \%) on LibriSpeech \textit{test-clean} under different online latency settings: (a) varying chunk size with no look-ahead ($N_l=0$) and (b) varying look-ahead size with a fixed 160~ms chunk size ($N_c=8$). The number of online registers and the number of OPC predicted future frames are fixed to $N_r=1$ and $N_f=4$, respectively.}
    \label{fig:latency_vs_wer}
\end{figure}

\noindent
\textbf{Latency vs. Accuracy.}
Figure~\ref{fig:latency_vs_wer} illustrates ASR performance on the LibriSpeech \textit{test-clean} set under various latency conditions. Figure~\ref{fig:1a} shows results with the lookahead fixed at $\Nl=0$ while varying the chunk size ($\Nc=8,16,32$). Notably, the performance gains are most pronounced at smaller chunk sizes, particularly at 160~ms latency ($\Nc=8$). Similarly, Figure~\ref{fig:1b} shows results with the chunk size fixed at $\Nc=8$ while varying the lookahead ($\Nl=0,2,4,8$), demonstrating that high accuracy is maintained even with zero lookahead. This indicates that OPC explicitly strengthens the mitigation of attention mismatch as discussed in Section~\ref{sec:mismatch}.

Notably, the combination of online registers and OPC achieves comparable accuracy at 160\,ms latency to the baseline operating at 320\,ms in Figure~\ref{fig:1a}. The same applies to the conditions with and without 40ms lookahead in Figure~\ref{fig:1b}. These results indicate that the proposed method can reduce latency while maintaining comparable recognition accuracy.

\noindent
\textbf{The Number of Future Frames.}
Table~\ref{tab:num-futures} presents the results obtained with different numbers of predicted future frames $\Nf$ in OPC. While predicting more future frames increases the difficulty of the auxiliary task, it may encourage the online registers to encode richer contextual information. This is expected to mitigate the gap of the online and offline self-attention shown in Figure~\ref{fig:attention}.

The results show that accuracy degrades when $\Nf$ is either too small ($\Nf=2$) or too large ($\Nf=8$). This suggests that long-term prediction is difficult and may negatively affect the learning process itself. The best performance was observed when $\Nf=4$, except in the offline condition on the \textit{test-other} set. Since the \textit{test-other} set contains more challenging samples for speech recognition, a richer future context may have been crucial.

\begin{table}[tbp]
    \centering
    \caption{Word error rate (WER, \%) of our model with Online Predictive Coding (OPC) for different numbers of predicted future frames ($N_f$), evaluated on LibriSpeech \textit{test-clean} and \textit{test-other}. All online decoding uses a \textbf{160~ms} chunk size ($N_c=8$) with no look-ahead ($N_l=0$).}
    \label{tab:num-futures}
    \begin{tabular}{@{}ccccc@{}}
        \toprule
        \multirow{2}{*}{\#frames $N_f$} & \multicolumn{2}{c}{test-clean} & \multicolumn{2}{c}{test-other} \\ \cmidrule(l){2-5} 
                                        & Offline       & Online         & Offline       & Online         \\ \midrule
        $2$                             & 2.68          & 3.49           & 6.48          & 9.83           \\
        $4$                             & \textbf{2.64} & \textbf{3.40}  & 6.41          & \textbf{9.65}  \\
        $6$                             & 2.65          & 3.43           & \textbf{6.39} & 9.72           \\
        $8$                             & 2.65          & 3.47           & 6.42          & 9.78           \\ \bottomrule
    \end{tabular}
\end{table}

\section{Conclusion}

In this paper, we proposed the pre-training framework named Online Predictive Coding (OPC), which explicitly encourages robust modeling of future context. By enabling online registers to encode predictive information about unseen future frames, our method reduces the offline–online attention gap. In addition, Dual-mode Layer Normalization mitigates distribution shifts between online and offline pathways, particularly those introduced by online registers. Our experiments showed that these modifications improve ASR performance, especially on more challenging test sets, without increasing algorithmic latency.

\section{Generative AI Use Disclosure}

Generative AI tools are used for grammar and style editing of the manuscript and for assistance in implementing experimental code. The tools were not used to write major parts of the manuscript, formulate the research questions, design the experiments, analyze the results, or draw conclusions. 

\bibliographystyle{IEEEtran}
\bibliography{mybib}

\end{document}